\documentclass[prl,letterpaper,twocolumn,showpacs,floatfix,preprintnumbers,superscriptaddress]{revtex4}
\usepackage{algorithm,algorithmicx,algpseudocode}
\usepackage{dcolumn,amsmath,amssymb,graphicx,color,placeins}
\usepackage{algorithmicx}
\usepackage{pdfpages}

\newcommand{\gras}[1]{\boldsymbol{#1}}

\begin{document}

\date{\today}

\title{Exact tensor hypercontraction: A universal technique for the resolution of
matrix elements of local, finite-range $N$-body potentials in many-body quantum
problems}

\author{Robert M. Parrish}
\affiliation{Center for Computational Molecular Science and Technology,\\
             School of Chemistry and Biochemistry,\\
             and School of Computational Science and Engineering,\\
             Georgia Institute of Technology, Atlanta, GA 30332-0400, United States}

\author{Edward G. Hohenstein}
\affiliation{Department of Chemistry and the PULSE Institute, Stanford University, Stanford, CA 94305}
\affiliation{SLAC National Accelerator Laboratory, Menlo Park, CA 94025}

\author{Nicolas F. Schunck}
\email{schunck1@llnl.gov}
\affiliation{Lawrence Livermore National Laboratory, Livermore, CA 94551}

\author{C. David Sherrill}
\email{sherrill@gatech.edu}
\affiliation{Center for Computational Molecular Science and Technology,\\
             School of Chemistry and Biochemistry,\\
             and School of Computational Science and Engineering,\\
             Georgia Institute of Technology, Atlanta, GA 30332-0400, United States}

\author{Todd J. Mart\'{\i}nez}
\email{toddjmartinez@gmail.com}
\affiliation{Department of Chemistry and the PULSE Institute, Stanford University, Stanford, CA 94305}
\affiliation{SLAC National Accelerator Laboratory, Menlo Park, CA 94025}

\begin{abstract}
Configuration-space matrix elements of $N$-body potentials arise naturally and
ubiquitously in the Ritz-Galerkin solution of many-body quantum problems. For
the common specialization of local, finite-range potentials, we develop the
eXact Tensor HyperContraction (X-THC) method, which provides a quantized
renormalization of the coordinate-space form of the $N$-body potential, allowing
for a highly separable tensor factorization of the configuration-space matrix
elements. This representation allows for substantial computational savings in
chemical, atomic, and nuclear physics simulations, particularly with respect to
difficult ``exchange-like'' contractions.
\end{abstract}


\maketitle



The physics of many-body quantum systems is often captured by local,
finite-range $N$-body potentials $\hat{V}(\gras{x}_1,\ldots, \gras{x}_N)$, where
$\gras{x}$ is any convenient parameterization of the physical space, e.g.,
position space $(\gras{x} \equiv \gras{r})$ or momentum space $(\gras{x} \equiv
\gras {k})$. Given some real, finite, one-particle Ritz-Galerkin basis set
$\{\psi_{i} (\gras{x}) \}$, the configuration-space representation of $\hat{V}$
is the integral tensor,
\begin{multline}
\langle i\ldots n|\hat V|i'\ldots n'\rangle
=
\int \mathrm{d}\gras{x}_1
\ldots
\int \mathrm{d}\gras{x}_N \\
\psi_{i} (\gras{x}_1)
\ldots
\psi_{n} (\gras{x}_N)
\hat V(\gras{x}_1, \ldots, \gras{x}_N)
\psi_{i'} (\gras{x}_1)
\ldots
\psi_{n'} (\gras{x}_N).
\label{eq:general_tensor}
\end{multline}
The generation, manipulation, and storage of this tensor is a major hurdle in
many-body quantum simulations. In order to overcome the computational difficulties
inherent to such high order tensors, it is common to introduce simplifying
approximations. For example, the Slater
approximation \cite{Slater:81:385} has been applied to reduce the numerical expense of
treating exchange terms involving the local, two-body Coulomb potential. Unfortunately,
such approximations can fail, as exemplified by the often spectacular self-interaction
errors induced by local approximations to exchange interactions \cite{Perdew:1982:1691}.
Another canonical example is nuclear density functional theory (DFT), where the need 
for computational savings is the main driver for the continued usage of energy density 
functionals (EDF) derived from the zero-range Skyrme-like pseudopotential 
\cite{Skyrme:1958:615}, in spite of severe problems at both two- and three-body levels 
\cite{Nozieres:Book,Blaizot:1976:435}. At the two-body level, even EDF derived from 
the finite-range Gogny pseudopotential \cite{Decharge:1980:1568}, which allows to avoid 
some of the limitations of Skyrme functionals \cite{Anguiano:2001:467}, contain the 
same phenomenological density-dependent terms recently shown to cause the collapse of 
all beyond mean-field methods 
\cite{Dobaczewski:76:054315,Duguet:79:044320,Lacroix:2009:044318,Bender:2009:044319}.
Removing density-dependences in the EDF, however, would probably require introducing 
explicit {\it finite-range} 3-body forces, which poses a serious computational challenge 
with current technology. It is thus clear that an improved algorithm for faithful and 
direct treatment of arbitrary local $N$-potentials (with $N\geq 2$) would be highly 
desirable .

In this Letter, we show that an \emph{exact} and \emph{separable} decomposition
exists for any local potential in a finite basis set built from polynomial
functions in any desired parameterization of the physical space. This decomposition
is motivated by our recently introduced Tensor HyperContraction (THC) method
for electronic
structure \cite{Hohenstein:2012:044103,Parrish:2012:LSTHC,Hohenstein:2012:METH},
which provided a phenomenological approximation for the electron repulsion
integrals involving the Coulomb potential in non-polynomial basis sets. The new
eXact Tensor HyperContraction (X-THC) representation
reveals two points of great importance for both electronic and nuclear structure problems.
First, THC approximation of the Coulomb interaction is exact for basis sets
which can be expressed in polynomial form (and thus the approximation in electronic
structure arises only because the basis functions used were of non-polynomial
form). Secondly, THC approximation is applicable not only to the two-body Coulomb
interaction but also to arbitrary local potentials commonly encountered in nuclear
structure (such as the Coulomb, Gogny, local forms of realistic three-body
potentials, etc.). Since the nuclear problem is already commonly formulated in
terms of polynomial basis sets, this implies that many problems in nuclear
structure can now be treated exactly with the \emph{lossless scaling
reduction} afforded by the X-THC representation. The first of these points
may aid markedly in the search for more efficient THC approximations in electronic
structure, while the second may yield unprecedented physical fidelity in nuclear
structure computations (especially within the context of nuclear DFT).

Below, we first demonstrate the key features of the X-THC representation through
the representative example of a one-dimensional, two-body problem in Cartesian
coordinates using Hermite functions. The $D$-dimensional, $N$-body
generalization of X-THC is then presented. Finally, we present an example
implementation of X-THC for the finite-range Gaussian potential in a basis of
Hermite functions, demonstrating that X-THC is both lossless and markedly efficient
in practice.

{\em X-THC Example - } Consider  a one-dimensional ($D=1$) problem in Cartesian
coordinates, involving a finite basis of $M+1$ Hermite functions $\{\psi_{i}
(x)\}$ (labeled from $0$ to $M$) with a local two-body ($N=2$) potential $\hat V
\equiv \hat V(x_1,x_2)$.  The potential matrix elements are,
\begin{multline}
\label{eq:2N}
\langle ij|\hat V|i'j'\rangle\equiv
\iint
\mathrm{d} x_1 \
\mathrm{d} x_2 \
\\
\psi_{i} (x_1)
\psi_{j} (x_2)
\hat{V}(x_1,x_2)
\psi_{i'} (x_1)
\psi_{j'} (x_2).
\end{multline}
The first stage in X-THC is to note that all $(M+1)^2$ products $ \psi_{i}
(x_1)\psi_{i'} (x_1)$ are exactly spanned by an orthonormal ``auxiliary'' basis
$\{\chi_A(x_1)\}$ consisting of $2M+1$ Hermite functions with a slightly
modified spatial range, $\chi_{A}(x_{1}) \equiv \psi_{A}(\sqrt{2}x_{1})$,
\begin{equation}
\psi_{i} (x_1)
\psi_{i'} (x_1)
=
\textstyle{\sum}_{A} [ii'A]
\chi_{A} (x_1),
\label{eq:linear_span}
\end{equation}
where,
\begin{equation}
\label{eq:3C}
[ii'A] \equiv
\int_{\mathbb{R}}
\mathrm{d} x_1 \
\psi_{i} (x_1)
\psi_{i'} (x_1)
\chi_{A} (x_1).
\end{equation}
This resolution is well known in the context of nuclear physics
\cite{Brody:1967:Tables,Talman:1970:273,Gogny:1975:399}, and is analogous to the
popular Density Fitting (DF) procedure of electronic structure theory
\cite{Whitten:1973:4496,Dunlap:1977:81,Vahtras:1993:514}.  In this context, the
decomposition is exact thanks to the closure properties of the polynomial-based
Hermite functions. The integrals are now given as,
\begin{equation}
\langle ij|\hat V|i'j'\rangle=
\sum_{AB} [ii'A]
[jj'B]
G^{AB}
,
\end{equation}
where,
\begin{equation}
\label{eq:2C}
G^{AB} \equiv
\iint_{\mathbb{R}^2}
\mathrm{d} x_1 \
\mathrm{d} x_2 \
\chi_{A} (x_1)
\chi_{B} (x_2)
\hat V(x_1,x_2).
\end{equation}
Thus, the fourth-order integral tensor is expressed as a product of second- and
third-order tensors. Even though we have compressed the fourth-order tensor,
this representation still precludes scaling reduction in ``exchange-like''
terms. A canonical example of such a term is the pairing field in
Hartree-Fock-Bogoliubov theory,
\begin{equation}
\Delta_{ij} \equiv
\sum_{i'j'} \langle ij|\hat V|i'j'\rangle \kappa_{i'j'}
=
\sum_{ABi'j'} [ii'A]
[jj'B]
G^{AB}
\kappa_{i'j'},
\end{equation}
where $\kappa$ is the pairing tensor. Despite the factorization, computing this
term still scales as ${\cal O}(M^4)={\cal O}(M^{2ND})$.

The critical step in THC is to resolve the three-index overlap integral
$[ii'A]$ to ``unpin'' the indices $i$ and $i'$ across some additional
linear-scaling index $P$. That is, we seek a decomposition of the form
$[ii'A] = \sum_{P} X_{i}^{P} X_{i'}^{P} Y_{A}^P$, where the range of $P$ is
$\mathcal O(M)$.  Thanks to the choice of a polynomial basis, the overlap
integral is exactly integrated by a $2M+1$-node Gaussian quadrature (in this
case, Gauss-Hermite) defined by the nodes and weights $\{ <x_P,w_P>\}$
\cite{AbramowitzStegun64}. Therefore, the quadrature grid index provides a
natural decomposition of the overlap integral,
\begin{equation}
[ii'A] = \sum_{P}
w_P
\psi_{i} (x_P)
\psi_{i'} (x_P)
\chi_{A} (x_P) =
\sum_{P} X_i^P
X_{i'}^P
Y_A^P,
\end{equation}
where  $X_i^P \equiv \psi_{i} (x_P)$ and $Y_A^P \equiv w_P \chi_{A} (x_p)$.
This is reminiscent of the discrete variable representation
\cite{Harris:1965:1515,Dickinson:1968:4209,Lill:1982:483,Baye:1986:2041} or
pseudospectral \cite{Friesner:1985:39} techniques of chemical physics. Defining
the intermediate $Z^{PQ} = \sum_{AB} Y_A^P G^{AB} Y_B^Q$, the full integral
(\ref{eq:2N}) is thus expressed as,
\begin{equation}
\langle ij | \hat V | i'j' \rangle =
\sum_{PQ} X_{i}^{P}
X_{j}^{Q}
Z^{PQ}
X_{i'}^{P}
X_{j'}^{Q}.
\end{equation}
This X-THC representation of the integral tensor is the key for the exact
${\cal O}(M^3)={\cal O}(M^{ND+1})$ treatment of the pairing term, via several
intermediate summations, indicated here by brackets for clarity,
\begin{multline}
\Delta_{ij} =
\sum_{PQi'j'} X_{i}^{P}
X_{j}^{Q}
Z^{PQ}
X_{i'}^{P}
X_{j'}^{Q}
\kappa_{i'j'} \\
 =
\sum_{P} X_{i}^{P}
\left[
\sum_{Q} X_{j}^{Q}
\left[
Z^{PQ}
\left[
\sum_{i'}X_{i'}^{P}
\left[
\sum_{j'}X_{j'}^{Q}
\kappa_{i'j'}
\right]
\right]
\right]
\right]
.
\end{multline}

\begin{figure}[!ht]
\begin{center}
\includegraphics[width=8.25cm, trim = 3.35cm 10.35cm 3.75cm 10.75cm, clip]{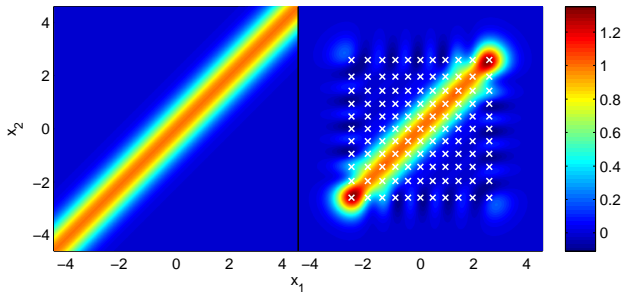}
\end{center}
\caption{(color online) Example of the X-THC process for a one-dimensional,
two-body Gaussian potential $\hat V(x_1,x_2) = \exp(-x_{12}^2)$ in Hermite
functions $\{\psi_{i} (x)\}$ up to $M=5$. Left: raw $\hat V(x_1,x_2)$. Right:
renormalized, quantizable $\bar V(x_1,x_2)$. White x's indicate the collocation
locations of the Gauss-Hermite quadrature to the quantized operator $\tilde
V(x_1,x_2)$.}
\label{fig:Z}
\vspace{-0.0in}
\end{figure}

{\em Interpretation - } At first glance, the $Z$ operator is a mere mathematical
intermediate, but there exists  a much richer interpretation: it is a quantized
renormalization of the coordinate-space representation of the potential operator
$\hat{V}$. To see this, we first consider the continuous, renormalized potential
operator $\bar{V}$, defined as,
\begin{equation}
\bar V(x_1,x_2) \equiv \sum_{AB} \chi_{A} (x_1) \chi_{B} (x_2) G^{AB}.
\end{equation}
This operator is not equivalent to the original in physical space, i.e., $\hat
V(x_1,x_2) \neq \bar V(x_1,x_2)$, yet the matrix elements of both operators are
identical, i.e., $\langle ij | \hat V | i'j' \rangle = \langle ij | \bar V |
i'j'\rangle$. The renormalized operator is simply the raw operator $\hat V$ with
all components outside of the finite product space $\{\psi_{i} (x_1) \psi_{i'}
(x_1)\} \Leftrightarrow \{\chi_{A} (x_1)\}$ projected out in each coordinate.
This projection is serendipitous: the coordinate-space integrand involving $\bar
V$ and the products of basis functions are exactly resolved by the Gaussian
quadrature for the auxiliary basis, while the corresponding integrand for $\hat
V$ is not exact under any finite quadrature due to the presence of ``alias''
components outside of $\{\psi_{i} (x_1) \psi_{i'} (x_1)\}$.  Applying the
Gaussian quadrature, we can quantize the renormalized operator $\bar V$ to
produce the discrete operator $\tilde V$, adding quadrature weights to account
for the spatial contribution of each point,
\begin{equation}
\tilde V(x_1,x_2) \equiv
w_P w_Q
\delta (x_1-x_P)
\delta (x_2-x_Q)
\bar V(x_1,x_2).
\end{equation}
As with $\bar V$, the matrix elements of $\tilde V$ are identical to those of
$\hat V$. Integrating $\tilde V$ instead of $\hat V$ naturally exposes the X-THC
factorization,
\begin{multline}
\langle ij | \hat V | i'j' \rangle
=
\langle ij | \tilde V | i'j' \rangle \\
=
\iint
\mathrm{d} x_1 \
\mathrm{d} x_2 \
\psi_{i} (x_1)
\psi_{j} (x_2)
\tilde V(x_1,x_2)
\psi_{i'} (x_1)
\psi_{j'} (x_2) \\
=
\sum_{PQ} X_{i}^{P}
X_{j}^{Q}
Z^{PQ}
X_{i'}^{P}
X_{j'}^{Q}.
\end{multline}
Here, the elements $Z^{PQ}$ are simply the quantized values of the renormalized
potential, with the weights rolled in, i.e., $Z^{PQ} = w_P w_Q \bar V(x_P,x_Q)$.
An example involving a Gaussian potential in Hermite functions is shown in
Figure \ref{fig:Z}. The renormalized potential (right) clearly shows the effects
of projection from the raw potential (left). The locations of the quantization
to $Z^{PQ}$ (the positions at which $\bar V$ can be discretized in a lossless
manner) are indicated with small white x's on the right.

This understanding of the $Z$ operator reveals that while X-THC is built from DF
and DVR techniques, the resultant supersedes both of the originals. In the
context of local potentials and polynomial basis sets, DF is always exact, but
does not provide separability of the $i$ and $i'$ indices, precluding scaling
reductions.  DVR techniques do provide separability, but are only exact when an
infinite quadrature is used, for an arbitrary choice of local potential. By
contrast, X-THC's particular merger of DF and DVR yields a perfect dealiasing
renormalization within a finite quadrature, providing a decomposition that is
both exact and separable for an arbitrary choice of local potential.


{\em Generalized X-THC - } The generalization of the one-dimensional, two-body,
Hermite function example above to $N$-body potentials in $D$-dimensions and
other choices of polynomial direct-product bases is straightforward. For X-THC
to hold, the one-particle basis must be of the $D$-dimensional
direct-product polynomial type, i.e., $\psi_{i} (\gras r) \equiv \prod_{\mu =
1}^{D} P_{i_{\mu}} (r_\mu) v_{\mu} (r_\mu) $.  In each dimension $\mu$,
$P_{i_{\mu}}$ is a polynomial of up to degree $i_{\mu}$, and $v_{\mu}$ is an
arbitrary weight function (analogous to the Gaussian term in the Hermite
functions above). Such basis sets are widely used in atomic and nuclear
many-body physics in various coordinate systems. Use of a direct-product
polynomial basis automatically guarantees closure: for the $M_\mu+1$ functions
in the $\mu^{\text{th}}$ dimension, the span $<\psi_{i_\mu} (
r_{\mu})\psi_{i_\mu'} (r_{\mu})>$ lies wholly inside a $2M_{\mu}+1$-function
auxiliary basis, defined by a set of polynomials orthogonal with respect to the
weight $|v_{\mu} (r_\mu)|^4$. Additionally, all quadratic products of auxiliary
functions are exactly integrated by a $2M_\mu+1$-node Gaussian quadrature
$\{<r_{P_\mu}, w_{P_\mu}>\}$ which can always be found, e.g., by the
Golub-Welsch algorithm \cite{GolubWelsch:1969}.

These properties allow for the X-THC factorization,
\begin{multline}
\label{eq:x-thc}
\langle i\ldots n |\hat V|i'\ldots n'\rangle = \\
\sum_{P\ldots W} X_i^{P}
\ldots
X_n^{W}
Z^{P\ldots W}
X_{i'}^{P}
\ldots
X_{n'}^{W},
\end{multline}
with each $X_{i}^{P}$ being the direct product of the $D$ underlying
$X_{i_\mu}^{P_\mu}$. $Z^{P\ldots W}$ is the  generalization of $Z^{PQ}$
to the case with $N$-body auxiliary integrals $G^{A\ldots N}$.

Within the X-THC representation, the representative generalization of
the pairing term, $\Delta_{i\ldots n} \equiv \langle i \ldots n | \hat V | i'
\ldots n' \rangle \kappa_{i'\ldots n'}$, now scales as ${\cal
O}(M_{\mu}^{ND+1})$, rather than ${\cal O}(M_{\mu}^{2ND})$, with no
approximation or restriction on the form of the local, finite-range potential
$\hat V$.

It is worth noting that common techniques to reduce the cost of treating
exchange-like terms involve approximating the potential to be direct-product
separable over $N_w$ terms, e.g., by approximating the Coulomb operator as a sum
of separable Gaussians \cite{Dobaczewski:2009:2361,Robledo:2010:044312}. This
reduces the conventional or DF cost of forming the generalized pairing tensor to
${\cal O} (M_{\mu}^{ND + N})$. X-THC can be applied to this approximate
separable potential, producing an ${\cal O}(M_{\mu}^{ND+1})$ implementation.
However, the separable form gives no particular scaling advantage in the
X-THC formalism, and can only reduce the prefactor and memory requirements.  A
more severe approximation is the invocation of a zero-range potential.  This is
typically formulated as a DVR-type quadrature in coordinate space, which can be
exact depending on the form of the zero-range operator
\cite{Dobaczewski:1997:166}. The asymptotic scaling of a pairing term involving
a zero-range potential is ${\cal O}(M_{\mu}^{ND+1})$, due to the first or last
transformation into or out of the grid index. Remarkably, this is the same
asymptotic scaling as X-THC. The zero-range potential will generally have lower
prefactor than X-THC (as there is only one grid coordinate in the zero-range
potential), but the asymptotic scalings are identical, and thus the tractability
limits should be comparable.  A summary of the scaling reductions afforded with
various factorization approaches and local potentials is shown in Table
\ref{tab:scaling}.

\begin{table}[!ht]
\caption{Computational scalings for the pairing term of an arbitrary local
potential in several approaches. $M_\mu$ is the order of the polynomial basis
in the  $\mu^{\mathrm{th}}$ degree of freedom, and the potential is $N$-body in $D$
dimensions. For simplicity, we consider the isotropic case where $M_\mu$ is the
same in all dimensions in this comparison. $N_w$ is the number of terms
retained in a separable approximation to the potential.  }
\label{tab:scaling}
\begin{center}
\begin{tabular}{lccc}
\hline \hline
Approach & General Local & Separable Local & Zero-Range \\
\hline
Conventional           & ${\cal O}(M_{\mu}^{2ND})$  & ${\cal O}(N_w
M_{\mu}^{ND+N})$ & $ {\cal O}(M_{\mu}^{ND+1})$\\
X-THC                  & ${\cal O}(M_{\mu}^{ND+1})$ & ${\cal O}(N_w
M_{\mu}^{ND+1})$ & $ {\cal O}(M_{\mu}^{ND+1})$\\
\hline \hline
\end{tabular}
\end{center}
\vspace{-0.25in}
\end{table}


{\em Practical Demonstration - } To illustrate the numerical equivalence and
practical utility of the X-THC approach, a hybrid MATLAB/C++ code was developed
to produce generalized pairing fields for $D$-dimensional, $N$-body forces in
Hermite functions. A complete description of the code is presented in the
supplemental material.

\begin{figure}[!ht]
\begin{center}
\includegraphics[width=8.25cm, trim = 6cm 8.75cm 6cm 9.0cm, clip]{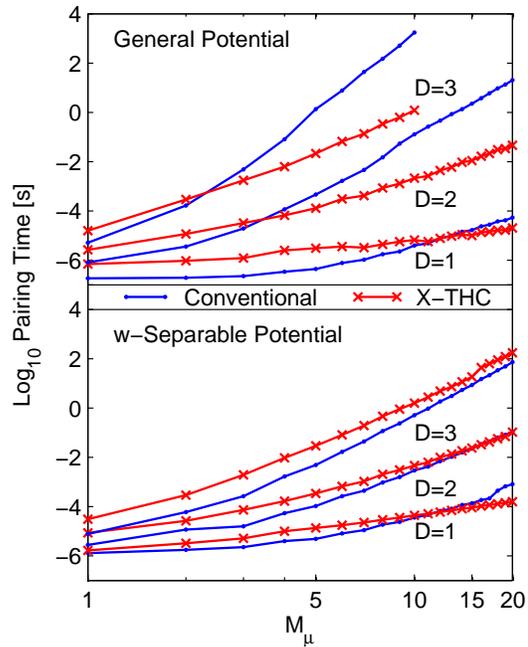}
\end{center}
\caption{
(color online) Wall times for pairing tensor formation as a function of
$M_{\mu}$ for $N=2$ (log-log scale). $N_w = 8$ for the separable potential.
}
\label{fig:timings}
\end{figure}

We have verified that the X-THC generalized pairing fields are exact within
machine precision (as expected mathematically). Figure \ref{fig:timings} shows
the computational gains which can be achieved from the X-THC factorization using
a representative example of $N=2$ and $D=1,2,3$.  For a general local potential,
X-THC is several orders of magnitude faster than conventional approaches for the
largest $M_\mu$ studied here. When the potential is written in separable
form, the X-THC scaling advantage is less dramatic, but X-THC becomes less
costly for the largest $M_\mu$ used in Figure \ref{fig:timings}. The X-THC
approach allows one to retain the general local potential and calculate the
{\em exact} pairing tensor in similar (or even less)\ computational effort as
with an {\em approximate} separable potential.

{\em Summary and Outlook - }
In this Letter, we have demonstrated that an exactly quantized renormalization
of any local, finite-range $N$-body potential exists in any situation where the
primary basis set may be composed of polynomial-based functions. This X-THC
representation provides for substantial computational scaling reduction of
contractions involving the local potential integral tensor, for instance by
reducing the formation of a representative generalization of the pairing tensor
from ${\cal O}(M_{\mu}^{2ND})$ to ${\cal O}(M_{\mu}^{ND+1})$.

In electronic structure, the concept of X-THC helps to codify and rationalize our
existing Least-Squares Tensor HyperContraction (LS-THC) approximation for
non-polynomial basis sets \cite{Parrish:2012:LSTHC}. The least squares procedure
introduced in that work actually
performs an implicit renormalization of the potential. Since the basis sets
used in our previous applications of LS-THC were not direct products of
polynomials (but rather atom-centered Gaussian functions), the decomposition
was necessarily an approximation. As the X-THC limit is approached, the
fidelity of the approximation will depend on
both the basis set resemblance to a set of polynomial-based
functions and the efficiency of the quadrature.
The physical picture provided by X-THC's explicit renormalization process
will almost certainly aid in the search for enhanced approximate THC recipes for
non-polynomial basis sets.

In nuclear structure, the potential applications for X-THC are immediate and
substantial. A crucial finding of this work is that the formal scaling of
operations involving arbitrary local potential operators is identical to that of
zero-range operators, without any loss in accuracy. This implies that the
finite-range two-body Gogny potentials of nuclear DFT can immediately be applied
with the same computational complexity as the
more approximate zero-range Skyrme
potentials. Tractability gains should be even more marked for three-body
potentials, paving the way for Hamiltonian-based nuclear energy densities derived
from effective, local, finite-range, density-independent, two- and three-nucleon pseudopotentials, which, by construction,
would be free of the current artifacts of nuclear DFT.


{\em Acknowledgments - } R.M.P.~is supported by a DOE Computational Science
Graduate Fellowship (Grant DE-FG02-97ER25308), particularly including a
practicum rotation at Lawrence Livermore National Laboratory. This material is
based on work supported by the National Science Foundation through grants to
C.D.S.~(Grant No.~CHE-1011360) and T.J.M.~(Grant No.~CHE-1047577), by the
Department of Defense through a grant to T.J.M.~(Office of the Director of
Defense Research and Engineering). It was performed under the auspices of
the US Department of Energy by the Lawrence Livermore National Laboratory
(Contract DE-AC52-07NA27344) through the Scientific Discovery through Advanced
Computing (SciDAC) program funded by U.S. Department of Energy, Office of
Science, Advanced Scientific Computing Research and Nuclear Physics. The Center
for Computational Molecular Science and Technology is funded through an NSF
CRIF award (Grant No.~CHE-0946869) and by Georgia Tech.


\bibliography{jrncodes,refs}
\bibliographystyle{aip}

\onecolumngrid
\newpage



\section*{Supplementary material}

# Supplemental material for: Exact tensor hypercontraction: A universal technique for the resolution of matrix elements of local, finite-range $N$-body potentials in many-body quantum problems

Robert M. Parrish,[1] Edward G. Hohenstein,[2, 3] Nicolas F. Schunck,[4, *] C. David Sherrill,[1, †] and Todd J. Martínez[2, 3, ‡]

[1]*Center for Computational Molecular Science and Technology,*
*School of Chemistry and Biochemistry,*
*and School of Computational Science and Engineering,*
*Georgia Institute of Technology, Atlanta, GA 30332-0400, United States*
[2]*Department of Chemistry and the PULSE Institute, Stanford University, Stanford, CA 94305*
[3]*SLAC National Accelerator Laboratory, Menlo Park, CA 94025*
[4]*Lawrence Livermore National Laboratory, Livermore, CA 94551*
(Dated: April 22, 2013)

## NOTATION

A rather large number of indices appear in the primary manuscript, so we summarize them here for clarity.

**Particle Number:** The particle number is denoted implicitly by the presence of ellipses in relevant equations. This index ranges from 1 to $N$.

**Dimension:** Each degree of freedom is indexed by $\mu$ and ranges from 1 to $D$ for each particle. In direct-product bases (e.g., X-THC), dimensionality plays a major role, as the total number of primary basis functions, auxiliary basis functions, and quadrature grid points scale as $\mathcal{O}(M_\mu^D)$ In non-direct-product bases (e.g., LS-THC), dimensionality has little meaning [2], so $D$ is usually assumed to be 1. In a non-direct-product basis, the number of primary basis functions, auxiliary basis functions, and quadrature grid points all generally scale as $\mathcal{O}(M)$.

**Primary Basis:** The primary single-particle basis is denoted by the indices $i$ to $n$ (bra), and $i'$ to $n'$ (ket). In a direct-product basis, $i$ is a composite index corresponding to the underlying direct-product of 1-dimensional primary basis functions, e.g., $|i\rangle \equiv |i_x\rangle|i_y\rangle|i_z\rangle$. In a polynomial direct-product basis, the 1-dimensional primary basis functions for the $\mu$th degree of freedom range from 0 to $M_\mu$ (the zero is a consequence of the polynomial definition of the basis). In a non-direct-product basis, the basis function indices range from 1 to $M$.

**Auxiliary Basis:** Auxiliary basis indices are denoted by the indices $A, B, \ldots$. In a direct-product basis, $A$ is a composite index corresponding to the underlying direct-product of 1-dimensional auxiliary basis functions, e.g., $|A\rangle \equiv |A_x\rangle|A_y\rangle|A_z\rangle$. In a polynomial direct-product basis, the 1-dimensional auxiliary basis functions in dimension $\mu$ range from 0 to $2M_\mu$ (the zero is a consequence of the polynomial definition of the basis). In a non-direct-product basis, the full auxiliary basis functions range from 1 to $M_A$, where increasing $M_A$ increases the fidelity of the approximate density fitting procedure.

**Quadrature Grid:** Quadrature grid indices are denoted by the indices $P, Q, \ldots$. In a direct-product basis, $P$ is a composite index corresponding to the union of the underlying one-dimensional quadratures, e.g., $\boldsymbol{r}_P \equiv (x_{P_x}, y_{P_y}, z_{P_z})$. In a polynomial direct-product basis, the indices for an exact quadrature for the $\mu$th degree of freedom range from 0 to $2M_\mu$ (the zero is a consequence of the polynomial definition of the basis). In a non-direct-product basis, the LS-THC quadrature grid contains $M_P$ points, where increasing $M_P$ increases the fidelity of the approximate LS-THC procedure.

For cases where two classes of indices are required to resolve a tensor element, a double subscript is used, e.g., the $P$-th grid point for the $\mu$th degree of freedom is denoted $r_{P_\mu}$. Also note that we use the generalized Einstein convention in this work: a repeated index on the right side of an equation is contracted over if it appears twice, or hypercontracted over if it appears more than twice, so long as the same index is not present on the left side of the same equation.

Note below that X-THC holds for any case in which the basis set is polynomial (in any representation of physical space, e.g., $\boldsymbol{x} = \boldsymbol{r}$), so long as the auxiliary potential integrals exist. This implies that a potential operator that is formulated in momentum space can be used in concert with a polynomial basis in position space, so long as the auxiliary integrals can be computed by any means available. An alternative way to picture this is that the momentum-space potential operator could be transformed to position-space via analytical Fourier transformation, without loss of locality. Having made this clarification, we will work entirely in the position space limit below.

## FULL $N$-BODY $D$-DIMENSIONAL X-THC

For clarity, we provide explicit definition of the full $N$-body $D$-dimensional X-THC representation here.

A direct-product basis founded on polynomials has the form,

$$\psi_i(\boldsymbol{r}) \equiv \prod_{\mu=1}^{D} P_{i_\mu}(r_\mu) v_\mu(r_\mu). \tag{1}$$

In each dimension $\mu$, $P_{i_\mu}$ is a polynomial of up to degree $i_\mu$, and $v_\mu$ is a weight function, typically chosen to bring the polynomial into $L_2(\mathbb{D})$, where $\mathbb{D}$ is the domain of the problem, and also to provide qualitative conformation to some *a priori* knowledge of the future shape of the wavefunction. The $i_\mu$ index ranges from 0 to $M_\mu$, so the polynomials range up to a maximum degree of $M_\mu$. Note that these polynomials do not have to be orthogonal, though they are often defined to be so. The product of polynomials being itself a polynomial of up to order $2M_\mu$, the local product $\psi_{i_\mu}^*(r_\mu)\psi_{i'_\mu}(r_\mu)$ in dimension $\mu$ lies inside the span of the $2M_\mu + 1$ auxiliary functions,

$$\chi_{A_\mu}(r_\mu) = \tilde{P}_{A_\mu}(r_\mu)|v_\mu(r_\mu)|^2, \tag{2}$$

where $\tilde{P}_{A_\mu}$ is a polynomial of up to order $2M_\mu$, often different from the primary basis polynomials $P_{i_\mu}$. We will choose these auxiliary functions to be orthonormal for convenience (this avoids DF metric matrices). Note that for a (generally complex) polynomial-based primary basis $\{\psi_{i_\mu}(r_\mu)\}$, we can always choose an exact, wholly real, orthonormal auxiliary basis $\{\chi_{A_\mu}(r_\mu)\}$ with $2M_\mu + 1$ functions. Thus, most of the complex conjugates on the auxiliary basis functions below are included only for convenience in the case that this work should need to be generalized to complex auxiliary bases.

The auxiliary functions $\chi_{A_\mu}(r_\mu)$ yield an exact DF representation for this basis,

$$\langle i \dots n|\hat{V}|i' \dots n'\rangle = [ii'A] \dots [nn'N]G^{A \dots N}. \tag{3}$$

The overlap integrals are separable in coordinates,

$$[ii'A] = \prod_{\mu=1}^{D} [i_\mu i'_\mu A_\mu]. \tag{4}$$

However, in general, the auxiliary potential integrals are *not* separable,

$$G^{A \dots N} \equiv \int \mathrm{d}\boldsymbol{r}_1 \ \dots \int \mathrm{d}\boldsymbol{r}_N$$
$$\chi_A^*(\boldsymbol{r}_1) \dots \chi_N^*(\boldsymbol{r}_N)\hat{V}(\boldsymbol{r}_1, \dots, \boldsymbol{r}_N). \tag{5}$$

To produce the THC representation, it remains to find an exact quadrature for the three-index overlap integrals. Owing to the closure in the product $i_\mu i'_\mu$, any quadrature which can exactly integrate the auxiliary overlap metric $[B_\mu A_\mu] = \delta_{B_\mu A_\mu}$ can exactly integrate the three-index overlap integrals $[i_\mu i'_\mu A_\mu]$, as the spans are identical in both cases. A quadrature which can exactly integrate all quadratic products of functions based on orthogonal polynomials of up to degree $2M_\mu$ is *precisely* the definition of the $2M_\mu + 1$-node Gaussian quadrature. Regardless of the choice of weight $v_\mu(r_\mu)$, the nodes and roots of this Gaussian quadrature can always be determined efficiently by the Golub-Welsch algorithm, giving the nodes and weights $\{<r_{P_\mu}, w_{P_\mu}>\}$ The overlap integral is then exactly resolved as,

$$[i_\mu i'_\mu A_\mu] = \underbrace{\psi_{i_\mu}^*(r_{P_\mu})}_{X_{i_\mu}^{*P_\mu}} \underbrace{\psi_{i'_\mu}(r_{P_\mu})}_{X_{i'_\mu}^{P_\mu}} \underbrace{\chi_{A_\mu}(r_{P_\mu}) w_{P_\mu}}_{Y_{A_\mu}^{P_\mu}}. \tag{6}$$

Note that the contraction index $P_\mu$ occurs not twice (as in a standard contraction operation) but three times (what we refer to as a "hypercontraction").

In the full direct-product basis, we will use the collapsed notation for the collocations of this Gaussian quadrature,

$$X_i^P = \prod_{\mu=1}^{D} X_{i_\mu}^{P_\mu}, \quad Y_A^P = \prod_{\mu=1}^{D} Y_{A_\mu}^{P_\mu}, \tag{7}$$

to save space. However, in real implementation, the direct-product separability of these two quantities is very important to achieve near-optimal scaling in formation of the X-THC factorization and subsequent utilization.

With these definitions, the X-THC $Z$ operator reads

$$Z^{P \dots W} = Y_A^P \dots Y_N^W G^{A \dots N}, \tag{8}$$

which yields the the full $N$-body $D$-dimensional X-THC,

$$\langle i \dots n|\hat{V}|i' \dots n\rangle = X_i^{*P} \dots X_n^{*W} Z^{P \dots W} X_{i'}^P \dots X_{n'}^W. \tag{9}$$

The $Z$ operator is not, in general, direct-product separable, but the factors $X$ and $Y$ are. In forming $Z$, the contraction of the $Y$ factors with the auxiliary potential integrals would naïvely scale as $\mathcal{O}(M_\mu^{ND+D})$, bearing in mind that the number of auxiliary functions and the number of quadrature points are both proportional to $\mathcal{O}(M_\mu^D)$. However, for each $Y$, we can perform the transformation in one dimension at a time (e.g., replacing $A_x$ with $P_x$, etc), reducing the formal scaling to $\mathcal{O}(M_\mu^{ND+1})$. Similarly, the separability of the $X$ and $Y$ factors is critically important to reduce the scaling of the generalized pairing term,

$$\Delta_{i \dots n} = \langle i \dots n|\hat{V}|i' \dots n'\rangle \kappa_{i' \dots n'}. \tag{10}$$

This contraction of a rank-$ND$ tensor with the integral tensor is in fact the worst possible scenario, as far as

the DF representation is concerned, since the compound contraction index involves *all* $N$ DF coefficient tensors,

$$\Delta_{i\ldots n} = d^A_{ii'} \ldots d^N_{nn'} G^{A\ldots N} \kappa_{i'\ldots n'}. \tag{11}$$

In general, this term will always scale as $\mathcal{O}(M^{2ND}_\mu)$ with both conventional and DF approaches, though the computational pre-factor is markedly higher in the latter case. When using X-THC, the generalized pairing term now reads,

$$\begin{aligned}
\Delta_{i\ldots n} &= \langle i\ldots n|\hat{V}|i'\ldots n'\rangle \kappa_{i'\ldots n'} \\
&= X^{*P}_i \ldots \left[X^{*W}_n \left[Z^{P\ldots W} \left[X^P_{i'} \ldots \left[X^W_{n'} \kappa_{i'\ldots n'}\right]\right]\right]\right]. \tag{12}
\end{aligned}$$

Any contraction involving $X$ here would naïvely scale as $\mathcal{O}(M^{ND+D}_\mu)$. However, for each $X$, we can perform the transformation in one dimension at a time (e.g., replacing $i'_x$ with $P_x$, etc), reducing the formal scaling to $\mathcal{O}(M^{ND+1}_\mu)$.

A common technique (usually an approximation) is to assert a $w$-separable form for the potential,

$$\hat{V}(\boldsymbol{r}_1,\ldots,\boldsymbol{r}_N) \equiv \sum_{w=1}^{N_w} \prod_{\mu=1}^D \hat{V}^w_\mu(r_{\mu_1},\ldots,r_{\mu_N}). \tag{13}$$

In this case, the integral tensor factors as,

$$\langle i\ldots n|\hat{V}|i'\ldots n'\rangle = \sum_{w=1}^{N_w} \prod_{\mu=1}^D \langle i_\mu\ldots n_\mu|\hat{V}^w_\mu|i'_\mu\ldots n'_\mu\rangle. \tag{14}$$

This allows the pairing tensor to be computed in $\mathcal{O}(N_w M^{ND+N}_\mu)$ via several intermediates. For X-THC, a $w$-separable potential allows for separability of the $Z$ operator, i.e.,

$$\begin{aligned}
\langle i\ldots n|\hat{V}|i'\ldots n'\rangle &= \sum_{w=1}^{N_w} \prod_{\mu=1}^D \\
&X^{*P_\mu}_{i_\mu} \ldots X^{*W_\mu}_{n_\mu} Z^{P_\mu\ldots W_\mu}_w X^{P_\mu}_{i'_\mu} \ldots X^{W_\mu}_{n'_\mu}. \tag{15}
\end{aligned}$$

This allows the pairing tensor to be computed in $\mathcal{O}(N_w M^{ND+1}_\mu)$. Note that this is the same formal scaling as X-THC in a general local potential, but the memory usage for $Z$ is lower, and the prefactor may or may not be lower, depending on how large $N_w$ is.

## DEMONSTRATION CODE

### Overview

To support the mathematical demonstration of exact X-THC resolution of the potential integral tensor, and to provide a practical example of the scaling gains provided by X-THC, a mixed MATLAB/C++ code was developed for the case of $D$-dimensional Hermite functions with $N$-body $w$-contracted Gaussian forces. In this code, the required potential integrals (in the primary or auxiliary basis) and possible X-THC factors are generated in MATLAB and written to disk, for the given basis size $M_\mu$, dimensionality $D$, number of bodies $N$, and number of $w$-contraction points $N_w$. For each integral technology and $w$-separable vs. $w$-nonseparable case, the generalized pairing field is computed for a randomly generated pairing tensor in a standalone C++ code for the particular $w$-separability and integrals technology case. In each C++ code, the integrals and factors are read in, the $w$ indices contracted over first if simulating a non-separable force, and then the generalized pairing tensor is computed according to the algorithms discussed below.

MATLAB was chosen for the integral and factor generation routines due to ease of implementation, and particular strength in treatment of arbitrary rank tensors. As this portion of the total procedure would typically be performed as a single-use overhead step (e.g., before HFB iterations or before application of the integrals in correlated methods), we do not include this step in the timings for the generalized pairing tensor, and thus there is no penalty for using the interpreted and rather memory-naïve MATLAB language for this stage. For the heavy linear algebra work of pairing tensor formation, C++ was selected for its "close to the metal" properties, particularly including explicit control of memory allocation and ability to swap pointers without performing explicit deep copy operations. Wherever possible (and for absolutely all contraction operations), BLAS calls are used, with the algorithms designed so as to allow for permutation of memory to be hidden in BLAS3 or BLAS2 operations as much as possible. The algorithms were formulated so as to rely preferentially on the BLAS3 DGEMM operation, followed by the BLAS2 DGEMV, followed by various BLAS1 operations. Every effort was expended to produce well-optimized algorithms, with the same amount of optimization present in both the X-THC and conventional methods. With these considerations, we believe that the timings reported are entirely representative of a practical application of the various integrals technologies, and show a wholly fair comparison of X-THC and conventional methods.

The C++ codes are compiled with the Intel `icpc 12.0.1` compiler, using `-O3` optimization. The BLAS calls are handled with Intel's very efficient `MKL 7.0.1` library, with threading disabled. Performance measurements are performed using the `PAPI 5.0.1` library, which features a wall timer accurate to $\sim 1$ microsecond. Accumulated averaging was performed to ensure that all pairing kernels ran for at least 1 second of wall time, which ameliorates startup and noise costs for small problem sizes. All timings were produced on a single-socket node featuring a quad-core 3.4 GHz Intel i7 Processor (Sandy Bridge) with 32 GB of DDR3 and 8 MB of L3 cache. All timings are for wall times using a single thread.

Note that we do not show results for density-fitted algorithms in this study, though we have coded generalized pairing routines using this approximation. All numerical results are essentially the same for conventional and X-THC approaches (e.g., numerically exact to within a small pre-factor of the machine epsilon). For $w$-separable forces with $N > 1$, the optimal pathway is to contract the DF intermediates to the conventional integrals, and then form the pairing tensor in the conventional manner. As a result, the DF timings for $w$-separable potentials are essentially indistinguishable from the conventional case. For non-$w$-separable potentials, the DF algorithms require the same $\mathcal{O}(M_\mu^{2ND})$ scaling as the conventional case, but the pre-factors are many orders of magnitude larger, due to the larger auxiliary basis sizes involved. Moreover, in the non-$w$-separable case, formation of the conventional integrals from the DF integrals exhibits a higher scaling of $\mathcal{O}(M_\mu^{2ND+1})$, and is therefore not a viable alternative. In any case, the conventional algorithm always outperforms the DF algorithm for the generalized pairing tensor, so we have elected to not include the DF results here. This failure of DF methods for "exchange-like" contractions is well known, and was the primary motivation for the development of the THC representation.

### Basis/Potential Choice

#### Hermite Function Primary Basis

The primary basis functions chosen for this demonstration are direct products of generalized Hermite functions,

$$\psi_i(\boldsymbol{r}) = \prod_\mu^D \psi_{i_\mu}(r_\mu). \tag{16}$$

Below, we will drop the $\mu$ labels and work in 1D ($\boldsymbol{r} \equiv x$) unless otherwise noted. The 1D generalized Hermite function is,

$$\psi_i(x) = (b_\mu)^{1/2} \underbrace{(\sqrt{\pi}2^i i!)^{-1/2} H_i(z)\exp(-z^2/2)}_{\phi_i(z)}, \tag{17}$$

where $H_i$ is the $i$-th Hermite polynomial and the nondimensional coordinate is,

$$z = b_\mu x. \tag{18}$$

$i$ runs from 0 to $M_\mu$. Note that we reserve $\phi_i(z)$ for the true non-dimensional Hermite function, where $b_\mu = 1$.

The auxiliary functions for this problem are also generalized Hermite functions,

$$\chi_A(x) = (\sqrt{2}b_\mu)^{1/2}(\sqrt{\pi}2^A A!)^{-1/2} H_A(z')\exp(-z'^2/2), \tag{19}$$

where now,

$$z' = \sqrt{2}b_\mu x, \tag{20}$$

and $A$ runs from 0 to $2M_\mu$. The THC quadrature for this problem is thus the $2M_\mu + 1$-node Gauss-Hermite quadrature with a spatial length scale of $\sqrt{2}b_\mu$.

#### Gaussian Potential

For flexibility, we use for the potential a linear combination of Gaussians. Given the set $\{< \alpha_w, \beta_w >\}$, the form of the potential is,

$$\hat{V}(x_1, \ldots, x_N) = \sum_w^{N_w} \sum_{\eta=1}^{N-1} \sum_{\xi=\eta+1}^{N} \alpha_w^{1/D} \exp(-\beta_w x_{\eta\xi}^2). \tag{21}$$

For a 2-body potential in 1 dimension, with $N_w = 1$ this reduces to the usual Gaussian force,

$$\hat{V}(x_1, x_2) = \alpha_w^{1/D} \exp(-\beta_w x_{12}^2). \tag{22}$$

For higher $N$, this is simply a sum of two-body Gaussian forces over all possible pairs of two-body coordinates, $x_{\eta\xi} = x_\eta - x_\xi$.

We use a set of $\{< \alpha_w, \beta_w >\}$ with $N_w = 8$ which approximates the Coulomb operator $1/r_{12}$ for all computations shown in this work. This allows us to show separate accuracy and timings results for the $w$-separable and non-$w$-separable cases, depending on whether we choose to sum over the $w$ index first or last.

### Potential Integrals (MATLAB)

All of the $D$-dimensional $N$-body potential integrals above can be constructed from primitive 1-dimensional 2-body potential integrals. For conventional integral approaches, four-index integrals in the primary basis are required. Within X-THC, two-center integrals in the auxiliary basis are required. Integrals of this type are discussed extensively in [1], where Moshinski transformations and length-scale transformations are used to provide analytical conventional integrals (the generalization to auxiliary integrals is straightforward).

#### Conventional Integrals

Following [1], the 1-dimensional 2-body primary integrals are computed as,

$$\langle ij|\hat{V}^w|i'j'\rangle = \alpha_w^{1/D} \iint \mathrm{d}x_1 \mathrm{d}x_2 \; \psi_i(x_1)\psi_j(x_2) \times$$
$$\exp(-\beta_w x_{12}^2)\psi_{i'}(x_1)\psi_{j'}(x_2)$$
$$= \alpha_w^{1/D} D_{mn}^{Aa} D_{m'n'}^{Ab} D_{aa'}(\eta_w)D_{ba'}(\eta_w), \tag{23}$$

where all summations run from 0 to $2M_\mu$. $D_{mn}^{Aa}$ are the Moshinski transformation coefficients, and $D_{aa'}(\eta_w)$ are the length-scale transformation coefficients (see below). The length-scale change parameter is

$$\eta_w = \frac{1}{\sqrt{1 + 2\beta_w/b_\mu^2}}. \tag{24}$$

*Auxiliary Integrals*

By extension, the 1-dimensional 2-body auxiliary integrals are computed as,

$$
\begin{aligned}
G_w^{AB} &= \alpha_w^{1/D} \iint \mathrm{d}x_1 \mathrm{d}x_2 \; \chi_A(x_1) \exp(-\beta_w x_{12}^2)\chi_B(x_2) \\
&= \frac{1}{\sqrt{2}b_\mu}\frac{\alpha_w^{1/D}}{\sqrt[4]{1 + 2\beta_w/b_\mu^2}} D_{AB}^{Nn} D_{nn'}(\eta_w) I_N I_{n'}, \quad (25)
\end{aligned}
$$

where again all summations run from 0 to $2M_\mu$. The $I_N$ quantities are primitive integrals over single Hermite functions of unit length,

$$I_N = \int \mathrm{d}z \; \psi_N(z) = \begin{cases} \sqrt[4]{\pi}\sqrt{2}\dfrac{(N-1)!!}{\sqrt{N!}} & N \text{ even} \\ 0 & N \text{ odd} \end{cases} \tag{26}$$

*Moshinski Transformation Coefficients $D_{mn}^{Aa}$*

The Moshinski transformation coefficients relate products of Hermite functions in Eulerian coordinates $x_1$ and $x_2$ to corresponding Hermite functions in the Lagrangian coordinates $X$ and $x$, where,

$$X = \frac{1}{\sqrt{2}}[x_1 + x_2], \quad x = \frac{1}{\sqrt{2}}[x_1 - x_2], \tag{27}$$

The transformation is

$$\psi_{n_1}(x_1)\psi_{n_2}(x_2) = D_{n_1,n_2}^{N,n}\psi_N(X)\psi_n(x). \tag{28}$$

As the Gaussian potential is central (i.e., depends only on $x$, not $X$), invoking the Moshinksi transformation reduces the two-coordinate potential integrals to a separable product of one-coordinate integrals in $X$ and $x$. If $n_1$ and $n_2$ each range from 0 to $M_\mu$, $N$ and $n$ each range from 0 to $2M_\mu$. The well-known selection rule is $n_1 + n_2 = N + n$. A simple, explicit formula for these coefficients is [1]

$$
\begin{aligned}
D_{n_1,n_2}^{N,n} = \delta_{n_1+n_2,N+n}&\left(\frac{n_1!n_2!}{N!n!}\right)\left(\frac{1}{\sqrt{2}}\right)^{N+n} \times \\
&\sum_{i,j=0}^{i<N,j<n,i+j=n_1}\binom{N}{i}\binom{n}{j}(-1)^j. \quad (29)
\end{aligned}
$$

However, this formula is unstable for large values of $N + n$ due to the summation over alternating quantities which are both large. We have therefore derived stable recurrence relations for the Moshinksi coefficients. The recurrence relation in the first coordinate is,

$$D_{n_1+1,n_2}^{N,n} = \frac{1}{\sqrt{2}}\left[\sqrt{\frac{N}{n_1+1}}D_{n_1,n_2}^{N-1,n} + \sqrt{\frac{n}{n_1+1}}D_{n_1,n_2}^{N,n-1}\right]. \tag{30}$$

And the corresponding recurrence relation in the other coordinate is,

$$D_{n_1,n_2+1}^{N,n} = \frac{1}{\sqrt{2}}\left[\sqrt{\frac{N}{n_2+1}}D_{n_1,n_2}^{N-1,n} - \sqrt{\frac{n}{n_2+1}}D_{n_1,n_2}^{N,n-1}\right]. \tag{31}$$

where

$$D_{0,0}^{0,0} = 1. \tag{32}$$

For implementation purposes, any Moshinski transformation coefficient with a "negative" index can be taken to be zero.

*Length-Scale Transformation Coefficients $D_{nn'}(\eta_w)$*

The length-scale transformation coefficients provide the correspondence between Hermite polynomials of different length-scales,

$$H_n(z) = (\eta)^{-1/2}D_{nn'}(\eta)H_{n'}(z' = z/\eta) \tag{33}$$

Note that the transformation automatically renormalizes the polynomials. $n$ and $n'$ both run from 0 to $M_\mu$.]

A closed-form expression of these coefficients is [1]

$$D_{nn'}(\eta) = F_{n,n'}(-1)^{\frac{n-n'}{2}}\left(\frac{n!}{n'!}\right)^{1/2}\frac{\eta^{n'+1/2}(1-\eta^2)^{\frac{n-n'}{2}}}{2^{\frac{n-n'}{2}}\left(\frac{n-n'}{2}\right)!}, \tag{34}$$

on the condition that $n > n'$, and that the parity agrees,

$$F_{n,n'} = \frac{1}{2}[1 + (-1)^{n+n'}] = \begin{cases} 1, & n+n' \text{ even} \\ 0, & n+n' \text{ odd} \end{cases} \tag{35}$$

Because $\eta \leq 1$, this formula appears to be universally stable. However, the numerators and denominators above both involve divergent values, so a log-space formalism is required for explicit evaluation to prevent overflow. This may lose a few digits of precision, so we have elected to use a recurrence relation for these coefficients, which is easily derived. The recurrence relation is,

$$
\begin{aligned}
D_{n+1,n'} = \eta\sqrt{\frac{n'+1}{n+1}}D_{n,n'+1} &+ \eta\sqrt{\frac{n'}{n+1}}D_{n,n'-1} \\
&- \sqrt{\frac{n}{n+1}}D_{n-1,n'}. \quad (36)
\end{aligned}
$$

where

$$D_{0,0}(\eta) = \sqrt{\eta}. \qquad (37)$$

For implementation purposes, any length-scale transformation coefficient with a "negative" index can be taken to be zero.

### D-Dimensional, N-Body Integrals

The generalization to $N$-body integrals as described above is carried out at the 1-dimensional stage (before the product over $\mu$ is carried out), e.g., for primary-basis integrals in the 3-body case,

$$\langle ijk|\hat{V}^w|i'j'k'\rangle = \langle ij|\hat{V}^w|j'j'\rangle\delta_{kk'} \\ + \langle ik|\hat{V}^w|i'k'\rangle\delta_{jj'} + \langle jk|\hat{V}^w|j'k'\rangle\delta_{ii'}. \quad (38)$$

For auxiliary basis integrals, there is no delta function in the third coordinate, but rather a normalized primitive Hermite integral, i.e., the quantity $I_N$ defined in the auxiliary potential integrals. Thus, the auxiliary potential integral is,

$$G_w^{ABC} = G_w^{AB}I_C + G_w^{AC}I_B + G_w^{BC}I_A. \qquad (39)$$

In practice, the generalization to 3-body integrals is performed in MATLAB on the 1-dimensional integrals, before integrals are written to disk.

The generalization to non-$w$-separable $D$-dimensional potentials is carried out by summing over $w$, e.g., in the 2-dimensional, 2-body case,

$$\langle ij|\hat{V}|i'j'\rangle = \sum_w^{N_w}\langle i_x j_x|\hat{V}^w|i'_x j'_x\rangle\langle i_y j_y|\hat{V}^w|i'_y j'_y\rangle, \quad (40)$$

or,

$$G^{AB} = \sum_w^{N_w} G_w^{A_x B_x} G_w^{A_y B_y}. \qquad (41)$$

In practice, the production of the nonseparable $Z$ factors or conventional integrals is carried out in blocks in C++, to save memory. The formation of these integrals is not counted in the pairing timings, as infinite memory is assumed. The overhead for this is many orders of magnitude larger for the conventional case than the X-THC case, due to the larger size of the rank-$2ND$ conventional integral tensor.

## X-THC Factors (MATLAB)

To complete the X-THC factorization, the Gauss-Hermite quadrature nodes/weights and collocation matrices $X$ and $Y$ are required.

### Quadratures

The THC grid for this problem is the $2M_\mu + 1$ node Gauss-Hermite quadrature with the spatial range parameter $\sqrt{2}b_\mu$.

First, the non-dimensional quadrature is generated. The orthonormal-basis position operator $X_{PQ} = \langle P|\hat{x}|Q\rangle$ in the auxiliary Hermite functions is,

$$X_{PQ} = \langle P|\hat{x}|Q\rangle = \sqrt{\frac{P+1}{2}}\left(\delta_{P+1,Q} + \delta_{P,Q+1}\right). \quad (42)$$

This operator is symmetric tridiagonal, with zero diagonal. The eigendecomposition is formed,

$$X_{PQ} = Q_{PP'}x_{P'}Q_{QP'}. \qquad (43)$$

The eigenvalues are the nondimensional quadrature nodes. The weights are determined by first generating the moment vector,

$$v_P = \sqrt[4]{\pi}\delta_{P0}, \qquad (44)$$

and applying the diagonalizing transformation,

$$v_{P'} = Q_{PP'}v_P. \qquad (45)$$

The full weights are then given by

$$w_{P'} = v_{P'}^2 \exp(x_{P'}^2). \qquad (46)$$

The nodes and weights are then transformed to the problem domain, by

$$x_P = \frac{x_{P'}}{\sqrt{2}b_\mu}, \quad w_P = \frac{w_{P'}}{\sqrt{2}b_\mu}. \qquad (47)$$

### Collocation

The $X$ and $Y$ X-THC factors require collocations at the quadrature nodes. This is accomplished by efficient, stable recurrence relations for the Hermite functions.

The non-dimensional coordinate is first computed,

$$z = b_\mu x. \qquad (48)$$

The first two non-dimensional Hermite functions are computed explicitly,

$$\psi_0(z) = \pi^{-1/4}\exp(-z^2/2), \qquad (49)$$

and

$$\psi_1(z) = \sqrt{2}z\pi^{-1/4}\exp(-z^2/2) = \sqrt{2}z\psi_0(z). \quad (50)$$

The recurrence relation is then applied iteratively,

$$\psi_{i+1}(z) = \sqrt{\frac{2}{i+1}}\left[z\psi_i(z) - \sqrt{\frac{i}{2}}\psi_{i-1}(z)\right] \\ = \sqrt{\frac{2}{i+1}}z\psi_i(z) - \sqrt{\frac{i}{i+1}}\psi_{i-1}(z). \quad (51)$$

And finally the length-scale normalization is added,

$$\psi_i(z) = (b_\mu)^{1/2}\psi_i(z). \tag{52}$$

The $X$ factor is,

$$X_i^P = \psi_i(x_P) = (b_\mu)^{1/2}\psi_i(z = b_\mu x_P), \tag{53}$$

and the $Y$ factor is,

$$Y_A^P = w_P\chi_A(x_P) = w_P(\sqrt{2}b_\mu)^{1/2}\psi_A(z' = \sqrt{2}b_\mu x_P). \tag{54}$$

If desired, the DF 3-index overlap integrals can be generated exactly (within numerical precision) from the X-THC factors,

$$[ii'A] = X_i^P X_{i'}^P Y_A^P. \tag{55}$$

The X-THC $Z$ operators are immediately formed from the $Y$ factors and corresponding $G$ auxiliary potential integrals,

$$Z_w^{PQ} = Y_A^P Y_B^Q G_w^{AB}. \tag{56}$$

**Generalized Pairing Algorithms (C++/BLAS)**

Algorithms for the generalized pairing tensor with conventional or X-THC integrals, and general or $w$-separable potentials, are described below and depicted in Algorithms 1-4. In these algorithms, we use ellipses to denote arbitrary rank in $D$ or $N$, and we follow the convention that dimensions are striped as the slow superindex, and particles are striped as the fast superindex [3]. The tensors used in these algorithms are stored in practice as a collapsed single-dimensional array (a `double*`), regularly striped so that the left-most index is the fastest dimension and the right-most index is the slowest dimension (Fortran order, which allows for convenient application of BLAS operations). For key contraction operations, we group sets of neighboring indices to form the three superindices $i$ (result row or fast index), $j$ (result column or slow index) and $k$ (contraction index) for use in GEMM,

$$C_{ij} = A_{ik}B_{kj}. \tag{57}$$

Note the use of color to emphasize the choice of superindices. By changing the order of $A$ and $B$ and altering the GEMM transposition arguments, we can take $i$, $j$, and $k$ in any order in the factor tensors $A$ or $B$ (e.g., the contraction index could actually be the row dimension in $A$). However, striding is not permitted with BLAS3 operations, e.g., a tensor contraction of the form,

$$C_{ijk} = A_{il}B_{jlk}. \tag{58}$$

would require explicit transposition to $B_{ljk}$ or $B_{jkl}$ to be able to group the compound $jk$ index. Our algorithms

are designed to eliminate such explicit transposition as much as possible. However, no transposition-free algorithm exists for X-THC $w$-separable potentials in arbitrary $N$, see the discussion below.

Note that our discussions and codes all assume isotropic basis sets for simplicity (e.g., $M_x = M_y$), but X-THC is certainly not restricted to this. In general, X-THC can handle differing values of $M_x$ and $M_y$, and even different classes of basis functions in each dimension (e.g. cylindrical coordinates).

*Conventional Integrals, General Potential*

See Algorithm 1. This algorithm is quite simple, but extraordinarily expensive (this is why non-$w$-separable forces are rarely used in high $D$ or $N$ cases). The physicists' integral tensor is used in a simple matrix-vector product with the generalized pairing tensor, which is carried out by GEMV in $M_\mu^{2ND} \equiv \mathcal{O}(M_\mu^{2ND})$ operations.

In practice, such an algorithm will almost certainly be orders of magnitude more expensive than reported here. In our demonstration, we simulate infinite memory by forming blocks of the integrals prior to GEMM (i.e., an integral direct procedure). The integral generation costs $\mathcal{O}(M_\mu^{2ND})$ (in this case by contracting out the $w$ index) with a much larger prefactor than the GEMV itself. In a fully generic potential, the integrals would have to be generated explicitly, at possibly even higher cost. This integral formation is not counted in the wall times reported here, but would increase the practical wall time considerably in the usual case that the rank-$2ND$ integral tensor does not fit in core memory.

*X-THC Integrals, General Potential*

See Algorithm 2. This algorithm works by cyclically transforming from configuration space $i'_\mu$ to the quantized coordinate space $P_\mu$ by GEMM, scaling the coordinate-space pairing tensor by $Z$, and then performing another cyclic transformation to bring $P_\mu$ back to $i_\mu$ via GEMM. The cyclic transformations are written so that the fast index in the current buffer is contracted off, and the replacement index is placed at the slow index of the result buffer. The pointers for the result and current buffer are then swapped, and the next contraction index automatically appears in the fast index of the current buffer, without any need for explicit transposition.

Formally, $\mathcal{O}(M_\mu^{ND+1})$ operations are required for the cyclic permutations. The prefactor arises from the $\sim 2M_\mu$ size (per dimension) of the quadrature index, and from the fact that two cyclic permutations are required.

This algorithm currently requires essentially three buffers of size $2^{ND}M_\mu^{ND}$ ($T$, $U$, and $Z$). An efficient blocked or disk-based algorithm could be developed if this

becomes a bottleneck, with considerable performance gains expected over conventional disk-based or integral-direct algorithms with general potentials.

### Conventional Integrals, w-Separable Potential

See Algorithm 3. This algorithm works by cyclically applying the integrals for dimension $\mu$ via GEMM, for each $w$ point. Explicit transposition is avoided by placing the replacement index $(i \dots n)_\mu$ as the slow index of the result, allowing $(i' \dots n')_{\mu+1}$ to be exposed as the new fast index. The formal scaling of this algorithm is $\mathcal{O}(N_w M_\mu^{ND+N})$ operations, with remarkably small external overhead. The memory requirement is essentially 2 $M_\mu^{ND}$ buffers ($T$ and $U$). The small overhead, combined with the efficiency of the long $(i' \dots n')_\mu$ contraction index and small memory footprint explains much of the current success of $w$-separable potentials.

### X-THC Integrals, w-Separable Potential

See Algorithm 4. This algorithm works by, for each $w$ point and dimension $\mu$, cyclically forward transforming from $i'_\mu$ to $P_\mu$, applying the $Z_w^{(P \dots W)_\mu}$ operation, and then cyclically backtransforming from $P_\mu$ to $i_\mu$. The formal scaling is $\mathcal{O}(N_w M_\mu^{ND+1})$ FMA (fused multiply-add) operations. The required memory is essentially two $2^N M_\mu^{ND}$ buffers $T$ and $U$, a factor of $2^N$ more than the conventional $w$-separable algorithm.

Unfortunately, an explicit transposition is required for this algorithm for general $N$. The genesis of this requirement is that the cyclic permutation is not carried through all $ND$ coordinates, but only $N$ coordinates at a time.

In the 2-body case, a specialized algorithm can be applied to avoid the explicit transposition. The transformation in each dimension reads,

$$
\begin{aligned}
U_{j'j'L} &\leftarrow T_{i'j'L} X_{i'}^P \\
U_{QLP} &\leftarrow T_{j'LP} X_{j'}^Q \\
U_{QLP} &\leftarrow T_{QLP} Z_w^{PQ} \\
U_{QLi} &\leftarrow T_{QLP} X_i^P \\
U_{Lij} &\leftarrow T_{QLi} X_j^Q
\end{aligned}
\tag{59}
$$

Here $L$ is the compound index corresponding to the other dimensions, and pointer swap is assumed between each step.

One additional modification can attenuate the prefactor somewhat, at the cost of doubling the buffer space. The forward transformation of the first dimension from $\kappa_{(i'j')_1 L}$ to $U_{L(PQ)_1}$ is the same for all $w$ points, as $\kappa$

is $w$-independent. Moreover, the back transformation of the last dimension from $U_{L(PQ)_D}$ to $\Delta_{L(ij)_D}$ does not depend on the individual $w$ points, but only on their sum. Thus the buffers $U_{L(PQ)_1}$ and $U_{L(PQ)_D}$ can be used in a prelude/epilogue construct. In the limit that $N_w \to \infty$, the savings are 100%, 50%, and 33%, for D = 1, 2,, and 3, respectively. In the limit that $N_w = 1$ or $D \to \infty$, the savings approach 0%, so the deployment of this modification would depend greatly on the context and memory capacity.

### Computational Results

Timings and accuracy results for generalized pairing tensor formation are shown in Figures 1 and 2, respectively, for various integral technologies, dimensions, number of bodies, and problem sizes. To help crystallize the information in the timing data, asymptotic scaling and predicted crossover metrics are presented in Tables I and II, respectively.

The accuracy results of Figure 2 depict the relative maximum residual $R$ in the pairing tensor, defined as,

$$
R_{\text{Method}} = \frac{\left\| \Delta_{i \dots n}^{\text{Method}} - \Delta_{i \dots n}^{\text{Reference}} \right\|_\infty}{\left\| \Delta_{i \dots n}^{\text{Reference}} \right\|_\infty}.
\tag{60}
$$

Here the $w$-separable conventional integral technology was selected as a reference for the accuracy [4]. It is apparent that all methods (conventional and THC) are exact to within a reasonable growth factor against the machine epsilon. In fact, the worst relative maximum residual seen here is less than $10^{-12}$ (compared to the double-precision machine epsilon of $2.2 \times 10^{-16}$), and does not seem to be grow markedly with respect to problem size. This provides strong numerical evidence that X-THC is a lossless compression of the potential integral tensor. Further agreement could doubtless be obtained if the Moshinksi coefficients were evaluated with higher precision [5].

The timing results of Figure 1 are quite clean as timings go, with very smooth increase with respect to problem size, indicating that the accumulation procedure (smaller $M_\mu$) or sheer size of the problem (larger $M_\mu$) are sufficient to eliminate the bulk of the noise that often plagues timing studies. On a log-log scale, all timing curves exhibit a slight upward concavity which quickly tends toward linearity for larger $M_\mu$ as the rate-limiting DGEMM-based steps become dominant relative to the lower-scaling operations. Two noticeable "jumps" exist: the last few points of the 2-body, 1-dimensional conventional separable case and the last few points of the 2-body, 3-dimensional THC separable case. These jumps are likely due to working set size limits exceeding some discrete performance threshold in hardware, for instance, cache or memory bank limits, respectively. From these

plots, power-law regression of the form,

$$t_{\text{Pairing}} = \alpha M_\mu^\beta, \tag{61}$$

is performed, using points selected above an $M_\mu$ which appears to be visually free of noise in each case. The $\beta$ from these regressions are depicted in Table I. The predicted THC crossover points from these regressions (the critical $M_\mu$ at which X-THC becomes practically superior to conventional integrals) are shown in Table II. For all cases in which an explicit crossover occurs, the predicted crossover point from the power-law model is within 1 $M_\mu$ of the observed value. Note that the observed asymptotic scaling is often less than the theoretical value. There are two geneses for this: residual contributions from lower scaling operations (which drags the scaling down at the cost of prefactor) and better GEMM/GEMV efficiency for larger matrix sizes (which makes GEMM/GEMV appear to scale better than cubic/quadratic as the matrix size increases). However, the relative scaling relationships between all integral technologies is retained.

From this point forward, it is useful to consider general and $w$-separable potentials separately.

For all cases of $N$ and $D$ in general local potentials, X-THC is markedly more efficient than conventional approaches. In all such cases, X-THC crosses over conventional (often at very small $M_\mu$ for $D > 1$), and is several orders of magnitude faster for the largest cases shown with $D > 1$. This is strong evidence for the immediate application of X-THC to problems involving general potentials.

For $w$-separable cases, X-THC always exhibits lower asymptotic scaling than conventional, and provides crossovers and sometimes significant speedups for $D = 1$ and $D = 2$. However, the narrower asymptotic separation between conventional and X-THC ($\beta_{\text{Conv}-\text{Sep}} - \beta_{\text{THC}-\text{Sep}} = N - 1$ vs. $\beta_{\text{Conv}-\text{Gen}} - \beta_{\text{THC}-\text{Gen}} = ND - 1$) exposes the prefactor of X-THC, particularly for larger $D$. This prefactor has two geneses: the formal FMA prefactor due to successive substitution of primary basis indices for larger quadrature indices and the lower efficiency of X-THC DGEMM operations compared to conventional DGEMM operations. For a 3-body, 3-dimensional $w$-separable potential, the crossover seems likely to occur just outside of the memory-limited problem size explicitly shown in Figure 1. In fact, the predicted crossover point for this case is $M_\mu = 10.3$. The 2-body, 3-dimensional $w$-separable potential is somewhat more sinister: a "jump" in the timings curve caused by some hardware boundary causes the expected crossover point to increase to a practically unreachable value of $M_\mu = 1789$.

The inability of X-THC to provide a practical crossover for the 2-body, 3-dimensional $w$-separable potential is an indication that this technique is not a panacea. However, we point out that the utilization of a $w$-separable potential throughout the literature seems to stem from the lack of an X-THC representation for a general potential: the approximation of the potential as $w$-separable was required to provide a tractable numerical recipe. In this light, X-THC treatment of general potentials provides a practical alternative pathway that avoids approximation of the potential as $w$-separable. For all of the cases studied here, the general X-THC curve is within roughly an order of magnitude of the conventional separable curve (in some cases even faster!). Moreover, X-THC does provide some gains in $w$-separable potentials, particularly for larger $N$. On more formal grounds, the demonstrated lossless asymptotic reduction of a generalized pairing term in any local potential from $\mathcal{O}(M_\mu^{2ND})$ to $\mathcal{O}(M_\mu^{ND+1})$ is a useful result in and of itself, considering that simply storing the pairing tensor requires $\mathcal{O}(M_\mu^{ND})$.

---

* Electronic address: schunck1@llnl.gov
† Electronic address: sherrill@gatech.edu
‡ Electronic address: toddjmartinez@gmail.com

---

**Algorithm 1** Generalized pairing algorithm: conventional integrals, general potential.

---

1: **procedure** PAIRING_CONV_GEN($\langle(i\ldots n)_1\ldots(i\ldots n)_D|\hat{V}|(i'\ldots n')_1\ldots(i'\ldots n')_D\rangle, \kappa_{(i'\ldots n')_1\ldots(i'\ldots n')_D}$ )
2:     Allocate $\Delta_{(i\ldots n)_1\ldots(i\ldots n)_D}$                                                  ▷ Target (Typically Preallocated)
3:     $\Delta_{(i\ldots n)_1\ldots(i\ldots n)_D} = \langle(i\ldots n)_1\ldots(i\ldots n)_D|\hat{V}|(i'\ldots n')_1\ldots(i'\ldots n')_D\rangle\kappa_{(i'\ldots n')_1\ldots(i'\ldots n')_D}$      ▷ GEMV, $\mathcal{O}(M_\mu^{2ND})$
4:     **return** $\Delta_{(i\ldots n)_1\ldots(i\ldots n)_D}$
5: **end procedure**

---

**Algorithm 2** Generalized pairing algorithm: X-THC integrals, general potential.

---

1: **procedure** PAIRING_THC_GEN($X_{i_\mu}^{P_\mu}, Z^{(P\ldots W)_1\ldots(P\ldots W)_D}, \kappa_{(i'\ldots n')_1\ldots(i'\ldots n')_D}$)
2:     Allocate $\Delta_{(i'\ldots n')_1\ldots(i'\ldots n')_D}$                                        ▷ Target (Typically Preallocated)
3:     Allocate $T_{(P\ldots W)_1\ldots(P\ldots W)_D}$                                      ▷ Scratch Array (Typically Preallocated)
4:     Allocate $U_{(P\ldots W)_1\ldots(P\ldots W)_D}$                                     ▷ Scratch Array (Typically Preallocated)
5:     $T_{(i'\ldots n')_1\ldots(i'\ldots n')_D} = \kappa_{(i'\ldots n')_1\ldots(i'\ldots n')_D}$                                  ▷ Deep Copy
6:     **for all** $\mu \in [1, D]$ **do**                                         ▷ Start Cyclic Permutation in $\mu$ and $\eta$
7:         **for all** $\eta \in [1, N]$ **do**
8:                 $U_{(\ldots n')_\mu\ldots(P\ldots W)_{\mu-1}(P)_\mu} = X_{i'_\mu}{}^{P_\mu}T_{(i')_\mu\ldots(n')_\mu\ldots(P\ldots W)_{\mu-1}}$        ▷ GEMM, $\mathcal{O}(M_\mu^{ND+1})$
9:                 swap($T, U$)                                    ▷ Pointer Swap
10:         **end for**
11:     **end for**                                             ▷ End Cyclic Permutation in $\mu$ and $\eta$
12:     $T_{(P\ldots W)_1\ldots(P\ldots W)_D}* = Z^{(P\ldots W)_1\ldots(P\ldots W)_D}$                     ▷ Hadamard Product, $\mathcal{O}(M_\mu^{ND})$
13:     **for all** $\mu \in [1, D]$ **do**                                       ▷ Start Cyclic Permutation in $\mu$ and $\eta$
14:         **for all** $\eta \in [1, N]$ **do**
15:                 $U_{(\ldots W)_\mu\ldots(i\ldots n)_{\mu-1}(i)_\mu} = X_{i_\mu}{}^{P_\mu}T_{(P)_\mu(\ldots W')_\mu\ldots(i\ldots n)_{\mu-1}}$        ▷ GEMM, $\mathcal{O}(M_\mu^{ND+1})$
16:                 swap($T, U$)                                    ▷ Pointer Swap
17:         **end for**
18:     **end for**                                             ▷ End Cyclic Permutation in $\mu$ and $\eta$
19:     $\Delta_{(i\ldots n)_1\ldots(i\ldots n)_D} = T_{(i\ldots n)_1\ldots(i\ldots n)_D}$                           ▷ Deep Copy
20:     **return** $\Delta_{(i\ldots n)_1\ldots(i\ldots n)_D}$
21: **end procedure**

---

**Algorithm 3** Generalized pairing algorithm: conventional integrals, $w$-separable potential.

---

1: **procedure** PAIRING_CONV_SEP($\langle i \dots n | \hat{V} | i' \dots n' \rangle_\mu^w$, $\kappa_{(i' \dots n')_1 \dots (i' \dots n')_D}$)
2:      Allocate $\Delta_{(i \dots n)_1 \dots (i \dots n)_D} = 0$          ▷ Target (Typically Preallocated)
3:      Allocate $T_{(i' \dots n')_1 \dots (i' \dots n')_D}$          ▷ Scratch Array (Typically Preallocated)
4:      Allocate $U_{(i' \dots n')_1 \dots (i' \dots n')_D}$          ▷ Scratch Array (Typically Preallocated)
5:      **for all** $w \in [1, N_w]$ **do**
6:          $T^w_{(i' \dots n')_1 \dots (i' \dots n')_D} = \kappa_{(i' \dots n')_1 \dots (i' \dots n')_D}$          ▷ Deep Copy
7:          **for all** $\mu \in [1, D]$ **do**          ▷ Start Cyclic Permutation in $\mu$
8:              $U^w_{\dots (i \dots n)_{\mu-1} (i \dots n)_\mu} = \langle i \dots n | \hat{V} | i' \dots n' \rangle_\mu^w T^w_{(i' \dots n')_\mu \dots (i \dots n)_{\mu-1}}$          ▷ GEMM, $\mathcal{O}(N_w M_\mu^{ND+N})$
9:              swap($T^w, U^w$)          ▷ Pointer Swap
10:          **end for**          ▷ End Cyclic Permutation in $\mu$
11:          $\Delta_{(i \dots n)_1 \dots (i \dots n)_D} += T^w_{(i \dots n)_1 \dots (i \dots n)_D}$          ▷ Contribution from $w$
12:      **end for**
13:      **return** $\Delta_{(i \dots n)_1 \dots (i \dots n)_D}$
14: **end procedure**

---

**Algorithm 4** Generalized pairing algorithm: X-THC integrals, $w$-separable potential.

---

1: **procedure** PAIRING_THC_SEP($X_{i_\mu}^{P_\mu}$, $Z_w^{(P \dots W)_\mu}$, $\kappa_{(i' \dots n')_1 \dots (i' \dots n')_D}$)
2:      Allocate $\Delta_{(i' \dots n')_1 \dots (i' \dots n')_D}$          ▷ Target (Typically Preallocated)
3:      Allocate $T_{(P \dots W)_1 \dots (i \dots n)_D}$          ▷ Scratch Array (Typically Preallocated)
4:      Allocate $U_{(P \dots W)_1 \dots (i \dots n)_D}$          ▷ Scratch Array (Typically Preallocated)
5:      **for all** $w \in [1, N_w]$ **do**
6:          $T_{(i' \dots n')_1 \dots (i' \dots n')_D} = \kappa_{(i' \dots n')_1 \dots (i' \dots n')_D}$          ▷ Deep Copy
7:          **for all** $\mu \in [1, D]$ **do**          ▷ Start Cyclic Permutation in $\mu$
8:              **for all** $\eta \in [1, N]$ **do**
9:                  $U_{(\dots n)_\mu \dots (i \dots n)_{\mu-1} (P)_\mu} = X_{i'_\mu}{}^{P_\mu} T_{(i')_\mu (\dots n')_\mu \dots (i \dots n)_{\mu-1}}$          ▷ GEMM, $\mathcal{O}(N_w M_\mu^{ND+1})$
10:                  swap($T, U$)          ▷ Pointer Swap
11:              **end for**
12:              $T_{\dots (i \dots n)_{\mu-1} (P \dots W)_\mu} *= Z_w^{(P \dots W)_\mu}$          ▷ SCAL, $\mathcal{O}(N_w M_\mu^{ND})$
13:              **for all** $\eta \in [N, 1]$ **do**
14:                  $U_{(n)_\mu \dots (i \dots n)_{\mu-1} (P \dots)_\mu} = X_{n_\mu}{}^{W_\mu} T_{\dots (i \dots n)_{\mu-1} (P \dots)_\mu (W)_\mu}$          ▷ GEMM, $\mathcal{O}(N_w M_\mu^{ND+1})$
15:                  swap($T, U$)          ▷ Pointer Swap
16:              **end for**
17:              $U_{\dots (i \dots n)_{\mu-1} (i \dots n)_\mu} = T_{(i \dots n)_\mu \dots (i \dots n)_{\mu-1}}$          ▷ Explicit Transposition
18:              swap($T, U$)          ▷ Pointer Swap
19:          **end for**          ▷ End Cyclic Permutation in $\mu$
20:          $\Delta_{(i \dots n)_1 \dots (i \dots n)_D} += T_{(i \dots n)_1 \dots (i \dots n)_D}$          ▷ Contribution from $w$
21:      **end for**
22:      **return** $\Delta_{(i \dots n)_1 \dots (i \dots n)_D}$
23: **end procedure**

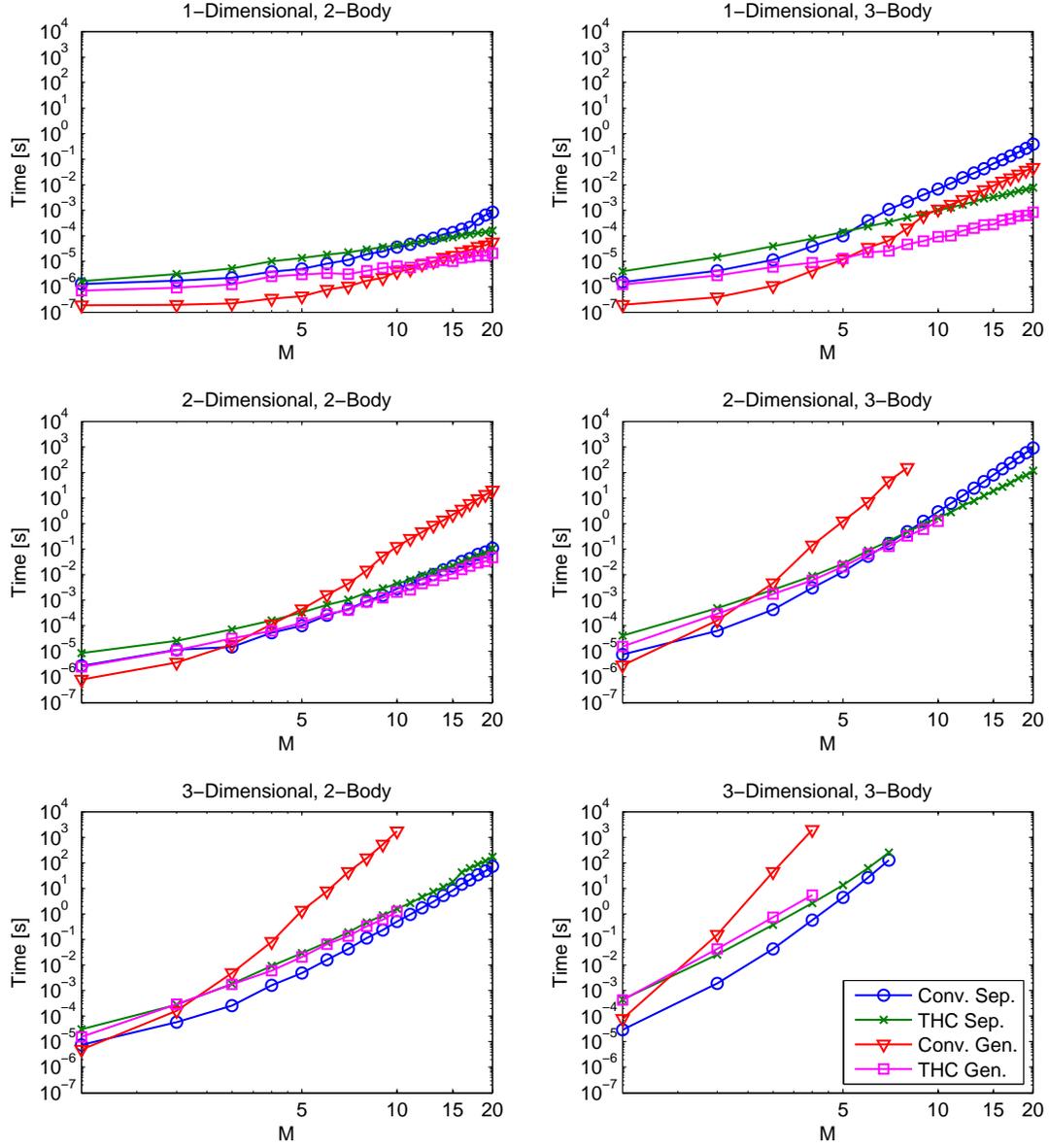

FIG. 1: Timings for generalized pairing tensor formation vs. problem size for various integral technologies, numbers of bodies $N$, and number of dimensions $D$.

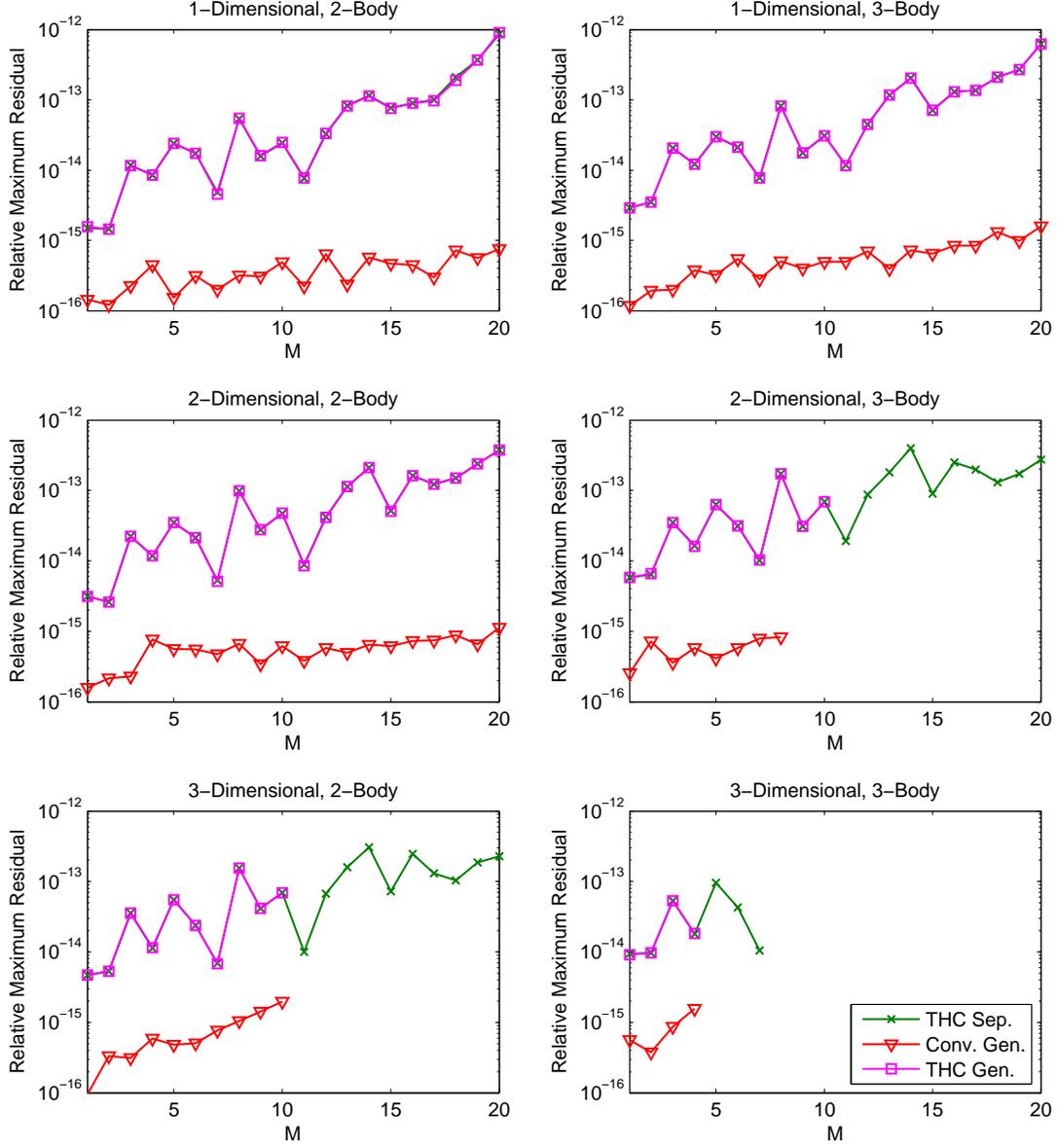

FIG. 2: Relative maximum residual for generalized pairing tensor formation vs. problem size for various integral technologies, numbers of bodies $N$, and number of dimensions $D$.

TABLE I: Asymptotic scalings of generalized pairing tensor formation with various integral technologies. Shown are observed scalings based on power-law fit to timing data and corresponding theoretical scaling (in parentheses).

| $D$ | $N$ | $\beta_{\mathrm{Conv-Sep}}$ | $\beta_{\mathrm{THC-Sep}}$ | $\beta_{\mathrm{Conv-Gen}}$ | $\beta_{\mathrm{THC-Gen}}$ |
|---|---|---|---|---|---|
| 1 | 2 | 4.5 ( 4) | 1.9 ( 3) | 3.9 ( 4) | 1.8 ( 3) |
| 1 | 3 | 5.8 ( 6) | 3.0 ( 4) | 5.5 ( 6) | 3.2 ( 4) |
| 2 | 2 | 5.3 ( 6) | 4.4 ( 5) | 7.3 ( 8) | 4.6 ( 5) |
| 2 | 3 | 8.3 ( 9) | 6.1 ( 7) | 10.5 (12) | 5.5 ( 7) |
| 3 | 2 | 7.2 ( 8) | 7.0 ( 7) | 10.3 (12) | 5.8 ( 7) |
| 3 | 3 | 8.9 (12) | 7.3 (10) | 13.7 (18) | 7.0 (10) |

TABLE II: Predicted THC vs. conventional crossover points for generalized pairing tensor formation based on power-law fit to timing data.

| $D$ | $N$ | $M_\mu^{\mathrm{Gen}}$ | $M_\mu^{\mathrm{Sep}}$ |
|---|---|---|---|
| 1 | 2 | 1.224E+01 | 1.172E+01 |
| 1 | 3 | 3.429E+00 | 5.170E+00 |
| 2 | 2 | 2.194E+00 | 1.657E+01 |
| 2 | 3 | 2.304E+00 | 7.928E+00 |
| 3 | 2 | 2.024E+00 | 1.789E+03 |
| 3 | 3 | 1.650E+00 | 1.035E+01 |



\end{document}